\documentclass[fleqn,a4paper,12pt]{article}
\usepackage{amssymb} 
\usepackage{amsbsy}
\usepackage{color}
\usepackage[pdftex]{hyperref}
\hypersetup{
  colorlinks=true,
  linkcolor = [rgb]{0,0,0.5},
  citecolor = [rgb]{0,0,0.5},
  urlcolor  = [rgb]{0,0,0.5}
}

\begin{document}

\counterwithin{equation}{section}
\def\theequation{\arabic{section}.\arabic{equation}}
\newtheorem{proposition}{Proposition}[section]


\newcommand\eref[1]{(\ref{#1})}
\newcommand\myvec[1]{\boldsymbol{#1}}
\newcommand\smallT{\scriptscriptstyle T}
\newcommand\mycev[1]{\boldsymbol{#1}^{\smallT}}

\newcommand\myvar[2]{\stackrel{\scriptscriptstyle #2}{#1}}
\newcommand\q[1]{\ifcase#1 q \or 
  \myvar{q}{1}\or 
  \myvar{r}{1}\or 
  \myvar{q}{2}\or 
  \myvar{r}{2}\or u\or v\else ? \fi}
\newcommand\z[1]{\ifcase#1 {\color{red}z} \or x \or {\color{red}x} \or t \or {\color{red}t} \else \stackrel{\scriptscriptstyle #1}{z} \fi}
\newcommand\myshifted[2]{\mathbb{E}_{#1}{#2}}

\title{\hbox to \textwidth{
  \bf\Large Solitons of the constrained Schr\"odinger equations.}}
\author{\hbox to \textwidth{Vekslerchik V.E.\hfill}}
\date{\parbox{0.95\textwidth}{\small%
  Usikov Institute for Radiophysics and Electronics, \\
  12, Proskura st., Kharkiv 61085,Ukraine  \\
  E-mail: \texttt{vekslerchik@yahoo.com}}}
\maketitle

\begin{abstract}
\noindent
We consider the linear vector Schr\"{o}dinger equation subjected to quadratic 
constraints. We demonstrate that the resulting nonlinear system is closely 
related to the Ablowitz-Ladik hierarchy and use this fact to derive the 
$N$-soliton solutions for the discussed model.
As an example of application of these results we present solitons 
of some vector nonlinear Schr\"{o}dinger equation with gradient nonlinearity.
\end{abstract}

\section{Introduction.}

In this paper we want to discuss some nonlinear and seemingly integrable model 
in which the nonlinearity arises from the imposed constraints. We follow the 
approach developed by Pohlmeyer who considered in \cite{P76} the \emph{linear} 
wave equation under \emph{quadratic} constraints. 
This approach, which has been generalized by various authors, leads to 
the so-called $\sigma$-models, which play an important role in 
modern mathematics and physics 
(see, e.g., the books \cite{K00,F13}). 

Here, we would like to find some solutions for the problem when 
similar constraints are applied to the linear Schr\"{o}dinger equation.
This problem is described by the action 
$\mathcal{S} = \int dx\, dt \, \mathcal{L}$ with the Lagrangian 
\begin{equation} 
\label{Lagrangian}
  \mathcal{L} = 
  i \myvec{\psi}^{\dagger} \myvec{\psi}_{t} 
  - \myvec{\psi}^{\dagger}_{x} \myvec{\psi}_{x} 
  + \lambda \left( \myvec{\psi}^{\dagger} \myvec{\psi} - 1 \right)
\end{equation} 
where $\myvec{\psi}$ is a two-dimensional complex vector 
which is a function of two real variables $t$ and $x$, 
$\myvec{\psi}=\myvec{\psi}(t,x)$,
$\myvec{\psi}^{\dagger}$ is its Hermitian conjugate and 
subscripts denote derivatives with respect to the corresponding variables.
The Lagrange multiplier $\lambda(t,x)$ is introduced to meet the constraint 
\begin{equation} 
\label{constraint}
  \myvec{\psi}^{\dagger} \myvec{\psi} = 1.
\end{equation} 
The subject of our study are the Euler–Lagrange equations for \eref{Lagrangian}, 
which can be written as 
\begin{equation} 
\label{eq:snlse}
\begin{array}{r} 
  i \myvec{\psi}_{t} 
  + \myvec{\psi}_{xx} 
  + \lambda \myvec{\psi}
  = 0, 
  \\
  - i \myvec{\psi}^{\dagger}_{t} 
  + \myvec{\psi}^{\dagger}_{xx} 
  + \lambda \myvec{\psi}^{\dagger}
  = 0 
\end{array} 
\end{equation} 
with 
\begin{equation}
\label{lambda}
  \lambda = 
  \mathop{\mbox{Im}} \myvec{\psi}^{\dagger} \myvec{\psi}_{t} 
  + \myvec{\psi}^{\dagger}_{x} \myvec{\psi}_{x}. 
\end{equation} 

The key point of this work is to demonstrate that equations 
\eref{eq:snlse} and \eref{lambda} can 
be `embedded' into the Ablowitz-Ladik hierarchy (ALH). 
In section \ref{sec:ALH} we show how one can obtain solutions for 
\eref{eq:snlse} from solutions for the equations of the ALH. 
Such approach was used in, for example, \cite{V94,V20} and was shown to be rather 
useful when one wants to find some particular solutions because the ALH is one 
of the most well-studied integrable systems. In section \ref{sec:solitons} we 
derive the $N$-soliton solutions for our problem by modification of the already 
known solitons of the ALH.
In section \ref{sec:gnlse} we consider an example of possible applications of 
the obtained results and present solitons of some vector nonlinear 
Schr\"{o}dinger equation (NLSE) with gradient nonlinearity.

\section{Ablowitz-Ladik hierarchy. \label{sec:ALH}}

The ALH was introduced in 1976 in \cite{AL76} as an infinite number of 
differential equations, the most famous of which is the discrete nonlinear 
Schr\"{o}dinger equation. 

Later, it has been reformulated as a system of a few functional equations 
generated by the Miwa shifts applied to the functions of an infinite number 
of arguments. 
The Miwa shifts, denoted by $\mathbb{E}_{\xi}$, are defined as 
\begin{equation}
  \mathbb{E}_{\xi} q( \boldsymbol{z} ) = q( \boldsymbol{z} + i [\xi] )
\end{equation}
where  
\begin{equation}
  q( \boldsymbol{z} ) = q( z_{1},z_{2},...) = q( z_{k})_{k=1,...,\infty}
\end{equation}
and
\begin{equation}
\label{def:Miwa}
  q( \boldsymbol{z} + i [\xi] ) 
  = 
  q( z_{1} + i\xi, z_{2} + i\xi^{2}/2,...) 
  = 
  q( z_{k} + i\xi^{k}/k )_{k=1,...,\infty}.
\end{equation}
In these terms, the ALH can be formulated as the following set of equations:
\begin{equation}
\label{alh:tau}
  \begin{array}{l} 
  0 = 
  \tau_{n} \myshifted{\xi}{\tau_{n}}
  - \tau_{n-1} \myshifted{\xi}{\tau_{n+1}}
  - \rho_{n} \myshifted{\xi}{\sigma_{n}}, 
  \\[2mm]  
  0 = 
  \tau_{n} \myshifted{\xi}{\sigma_{n}} 
  - \sigma_{n} \myshifted{\xi}{\tau_{n}}
  - \xi \tau_{n-1} \myshifted{\xi}{\sigma_{n+1}}, 
  \\[2mm]
  0 = 
  \rho_{n} \myshifted{\xi}{\tau_{n}}
  - \tau_{n} \myshifted{\xi}{\rho_{n}} 
  - \xi \rho_{n-1} \myshifted{\xi}{\tau_{n}}. 
  \end{array}
  \quad
  n \in \left( -\infty, ..., \infty \right).
\end{equation}
Strictly speaking, the above equations constitute only a half of the hierarchy, 
which is known to consist of two similar sub-hierarchies (the so-called 
`positive' and `negative' flows). However, for our current purposes, we may 
restrict ourselves to \eref{alh:tau}. 

Now, we will derive some consequences of \eref{alh:tau}, which we use below to 
solve our problem.

First we introduce, for a fixed value of $n$, 
\begin{equation} 
  n = 0,
\end{equation} 
four new functions, 
\begin{equation}
\label{def:q14}
  \q1 = \frac{\sigma_{0}}{\tau_{0}}, \qquad
  \q2 = \frac{\rho_{0}}{\tau_{0}},   \qquad
  \q3 = \frac{\tau_{1}}{\tau_{0}},   \qquad
  \q4 = \frac{\tau_{-1}}{\tau_{0}}.
\end{equation}
The original Ablowitz-Ladik equations were formulated in terms of $\q1$ and 
$\q2$ (with the $n$-dependence restored), 
and the first two equations in \eref{def:q14} are the standard way to 
introduce the tau-functions in order to arrive at {\em bilinear} equations 
\eref{alh:tau}. The second pair of functions, $\q3$ and $\q4$, usually was not 
considered (or even introduced) in the framework of the Ablowitz-Ladik equations. 
However, as we see in what follows, they are a `natural' complement to $\q1$ and 
$\q2$ and will play an essential role in this work. 

One can show that functions defined in \eref{def:q14} satisfy 
\begin{equation}
\label{shx:q1}
  \myshifted{\xi}{\q1} - \q1 = \xi \q4 \myshifted{\xi}{\q5}, \qquad
  \myshifted{\xi}{\q2} - \q2 = - \xi \q6 \, \myshifted{\xi}{\q3} 
\end{equation}
\begin{equation}
 \myshifted{\xi}{\q3} - \q3 = - \xi \q2 \myshifted{\xi}{\q5}, \qquad
 \myshifted{\xi}{\q4} - \q4 = \xi \q6 \, \myshifted{\xi}{\q1} 
\label{shx:q4}
\end{equation}
where 
\begin{equation}
\label{def:q56}
  \q5 = \frac{\sigma_{1}}{\tau_{0}}, \qquad
  \q6 = \frac{\rho_{-1}}{\tau_{0}}
\end{equation}
together with the constraint 
\begin{equation}
\label{uni:q}
  \q1 \q2 + \q3 \q4 = 1.
\end{equation}
Returning from the functional equations to the differential ones 
with variables $z_{1}$ and $z_{2}$ being replaced with $\z1$ and $\z3$, 
\begin{equation} 
  \z1 = z_{1}, \quad \z3 = z_{2},
\end{equation} 
one can show, by means of the expansion
\begin{equation}
\label{miwa:series}
  \mathbb{E}_{\xi} q  = 
  q
  + i \xi q_{\z1}
  + \frac{\xi^{2}}{2} 
    \left( i q_{\z3} - q_{\z1\z1} \right)
  + O\left( \xi^{3} \right),
\end{equation}
that functions $\q1$, $\q2$, $\q3$ and $\q4$ satisfy  
\begin{equation}
\label{dx:q1}
  i \q1_{\z1} = \q5 \q4, 
\qquad 
  i \q2_{\z1} = - \q6 \q3, 
\end{equation}
\begin{equation}
\label{dx:q4}
  i \q3_{\z1} = - \q5 \q2, 
\qquad 
  i \q4_{\z1} = \q6 \q1 
\end{equation}
and
\begin{equation}
\label{dt:q1}
 \q1_{\z3} = \q5_{\z1}\q4  - \q5 \q4_{\z1}, 
\qquad
 \q2_{\z3} = \q6_{\z1}\q3  - \q6 \q3_{\z1}, 
\end{equation}

\begin{equation}
\label{dt:q4}
  \q3_{\z3}  = \q5 \q2_{\z1} - \q5_{\z1} \q2, 
\qquad
  \q4_{\z3} = \q6 \q1_{\z1} - \q6_{\z1} \q1. 
\label{dt:q4}
\end{equation}
Now, we introduce two 2-vectors, 
\begin{equation}
\label{def:qr}
  \myvec{q}= (\q1, \; \q3 )^{\smallT}, 
  \qquad  
  \myvec{r}= ( \q2, \; \q4 )^{\smallT} 
\end{equation}
and rewrite the above equations in the vector form,
\begin{equation}
\label{eq:dxq}
  \myvec{q}_{\z1} = \q5 \, \mathsf{\sigma}_{2} \, \myvec{r},
\qquad
  \myvec{r}_{\z1} = - \q6 \, \mathsf{\sigma}_{2} \, \myvec{q}
\end{equation}
and 
\begin{equation}
  \myvec{q}_{\z3} = 
  i \q5 \q6 \myvec{q} 
  + i \q5_{\z1} \, \mathsf{\sigma}_{2} \, \myvec{r},
\qquad
  \myvec{r}_{\z3} 
  = 
  - i \q5 \q6 \, \myvec{r}
  + i \q6_{\z1} \, \mathsf{\sigma}_{2} \, \myvec{q},
\label{eq:dtr}
\end{equation}
where 
$\sigma_{2} = $
{\small $\left(\begin{array}{cc} 0 & -i \\ i & 0 \end{array}\right)$}. 

The restriction \eref{uni:q} now becomes 
\begin{equation}
\label{uni:qr}
 \mycev{r} \myvec{q} = 1.
\end{equation}
By a simple algebra one can obtain the following consequence of 
\eref{eq:dxq} and \eref{eq:dtr}: 
\begin{equation}
\label{nlse:q}
  i \myvec{q}_{\z3} 
  + \myvec{q}_{\z1\z1} 
  + \lambda \myvec{q}
  = 0,
\end{equation}
\begin{equation}
\label{nlse:r}
  - i \mycev{r}_{\z3} 
  + \mycev{r}_{\z1\z1} 
  + \lambda \mycev{r}
  = 0
\end{equation}
with 
\begin{equation} 
\label{lambdauv}
 \lambda = 2 \q5 \q6  
\end{equation} 
as well as the identities
\begin{equation}
\label{eq:lambda}
  \lambda = 
  - i \mycev{r} \myvec{q}_{\z3} 
  - \mycev{r}   \myvec{q}_{\z1\z1} 
  =  
   i \mycev{r}_{\z3} \myvec{q} 
  - \mycev{r}_{\z1\z1} \myvec{q}. 
\end{equation}

It is easy to see that equations \eref{nlse:q} and \eref{nlse:r} together with 
\eref{eq:lambda} are the Euler–Lagrange equations for the action 
$\mathcal{S} = \int d\z1\,d\z3 \; \mathcal{L}$ with the Lagrangian 
\begin{equation}
\label{lagr:qr}
  \mathcal{L} = 
  i \mycev{r} \myvec{q}_{\z3} 
  - \mycev{r}_{\z1} \myvec{q}_{\z1} 
  + \lambda \left( \mycev{r}\myvec{{q}} - 1 \right)
\end{equation}
describing the vector \emph{linear} Schr\"{o}dinger system under 
the \emph{bilinear} restriction \eref{uni:qr}.
In other words, we have demonstrated that starting from \eref{alh:tau} one 
can obtain solutions of the `two-field' version of our problem.

It should be noted that problem with the Lagrangian \eref{lagr:qr} is more 
general than one with \eref{Lagrangian} and our original problem is a reduction 
of the former: $\myvec{r}^{\smallT} = \myvec{q}^{\dagger}$. 
In physical applications, models involving $\myvec{\psi}$ and 
$\myvec{\psi}^{\dagger}$ appear more often than similar models with two 
independent fields, as $\myvec{q}$ and $\myvec{r}$ in our case. 
However it is a common practice even in the cases of problems with complex 
fields, like, for example, the NLSE, consider $\myvec{\psi}$ and 
$\myvec{\psi}^{\dagger}$ as distinct variables because most part 
(but surely not all) of the calculations depend on the algebraic structure of the 
equations and not on the involution 
$\myvec{\psi} \leftrightarrow \myvec{\psi}^{\dagger}$.

An interesting fact, which has no direct relevance to our problem, is 
that $\q5$ and $\q6$ defined in \eref{def:q56} satisfy 
\begin{equation}
\label{nlse:uv}
\left\{ \begin{array}{r} 
  i \q5_{\z3}
  + \q5_{\z1\z1}
  + 2 \q5^{2} \q6
  = 0, 
\\
  - i \q6_{\z3} 
  + \q6_{\z1\z1} 
  + 2 \q5 \q6^{2}
  = 0. 
  \end{array} \right.
\end{equation}
(we give a proof of this statement in Appendix A).
So, as a by-products, we have obtained solutions for the NLSE.
 
Till now, the correspondence between the ALH and the model \eref{lagr:qr} 
was rather general: \emph{any} solution for \eref{alh:tau} provides solution 
for \eref{nlse:q}--\eref{eq:lambda}. However, not all of them can be used 
to obtain solutions with $\myvec{q}$ and $\myvec{r}$ being related by  
$\myvec{r}^{\smallT} = \myvec{q}^{\dagger}$. 
One thus needs some additional work. 
Moreover, we have to make some slightly nonstandard steps. 
The case is that the `natural' involution for the ALH-like equations is
$\rho_{n} = \pm\sigma_{n}^{*}$, $\tau_{n} = \tau_{n}^{*}$ where $*$ stands 
for the complex conjugation. Clearly, such involution does not provide 
necessary relation between $\q3$ and $\q4$.
Nevertheless, this  issue can be resolved.
In the next section we construct $N$-soliton solutions for our problem by 
modifying the already known ones that have been derived earlier for the ALH 
and  discuss the issue of involution in more detail.

\section{$N$-soliton solutions. \label{sec:solitons}}

The main part of the structure of the soliton solutions are the so-called 
`soliton matrices', that satisfy the system of the Sylvester equations
\begin{equation}
\label{sol:sy}
  \mathsf{L} \mathsf{A} - \mathsf{A} \mathsf{R} 
  = | \alpha \rangle \langle a |, 
  \qquad
  \mathsf{R} \mathsf{B} - \mathsf{B} \mathsf{L} 
  = | \beta \rangle \langle b |
\end{equation}
and that have been repeatedly used in the framework of the Cauchy matrix 
approach (see, e.g., chapter 9 of  \cite{HJN16}). 

Here, $\mathsf{L}$ and $\mathsf{R}$ are constant diagonal complex matrices,
\begin{equation}
  \mathsf{L} = \mbox{diag}\left( L_{1}, ... , L_{N} \right), \qquad
  \mathsf{R} = \mbox{diag}\left( R_{1}, ... , R_{N} \right), 
\end{equation}
$| \alpha \rangle$ and $| \beta \rangle$ are constant $N$-columns,
\begin{equation}
  | \alpha \rangle = \left( \alpha_{1}, ... , \alpha_{N} \right)^{\smallT}, 
  \qquad
  | \beta \rangle = \left( \beta_{1}, ... , \beta_{N} \right)^{\smallT}, 
\end{equation}
while $N$-rows $\langle a |$ and $\langle b |$
\begin{equation}
  \langle a | = \left( a_{1}, ... , a_{N} \right), 
  \qquad
  \langle b | = \left( b_{1}, ... , b_{N} \right) 
\end{equation}
depend on the coordinates and, in turn, determine the coordinate dependence 
of the $N \times N$ matrices $\mathsf{A}$ and $\mathsf{B}$.

The recipe for soliton solutions consists of two parts. 
The first step is to `construct' functions $\q1$ ... $\q4$ of 
the matrices and vectors introduced above. 
As was mentioned in the Introduction, we use the already known results for the 
ALH. So, we follow section 2.2 of the paper \cite{V15}, and present functions 
\eref{def:q14} as 
\begin{equation}
\label{sol:q1}
 \q1 = 
 \langle a |  \mathsf{R}^{-1}  \mathsf{F} | \beta \rangle, 
\end{equation}
\begin{equation}
\label{sol:q2}
  \q2
  = 
  \langle b |  \mathsf{L}^{-1}  \mathsf{G} | \alpha \rangle, 
\end{equation}
\begin{equation}
\label{sol:q3}
  \q3 = 
  1 + \langle a | \mathsf{R}^{-1} \mathsf{F} \mathsf{B} | \alpha \rangle, 
\end{equation}
\begin{equation}
\label{sol:q4}
  \q4 = 
  1 + \langle b | \mathsf{L}^{-1} \mathsf{G} \mathsf{A} | \beta \rangle
\label{sol:q4}
\end{equation}
where 
\begin{equation}
\label{sol:FG}
  \mathsf{F} = \left( 1 +  \mathsf{B}  \mathsf{A} \right)^{-1}, \qquad
  \mathsf{G} = \left( 1 +  \mathsf{A}  \mathsf{B}\right)^{-1}. 
\end{equation}
The next step, is the dependence of $\mathsf{A}$, $\mathsf{B}$ \emph{etc} on the 
variables of the hierarchy $z_{1}$, $z_{2}$, ... 
(and, in particular, on $\z1$ and $\z3$). 
Again, we do not invent anything new and use the `classical' prescription.
The case is that usually, in soliton solutions, the $z_{k}$-dependence 
appears through various exponential functions like 
$\exp[ -i \phi(\boldsymbol{z}) ]$ with
$\phi(\boldsymbol{z}) = \sum_{k=1}^{\infty} c^{k} z_{k}$ 
(the simplest non-trivial series).
The importance of such functions stems from the identity 
\begin{equation} 
  \myshifted{\xi}{\phi} - \phi = i \sum_{k=1}^{\infty} \frac{(c\,\xi)^{k}}{k}
  = - i \ln( 1 - c\xi )
\end{equation} 
and, as a result, from the fact that 
$\exp[ -i \phi(\boldsymbol{z}) ]$ is an eigenfunction 
of the shift operator $\mathbb{E}_{\xi}$:
\begin{equation} 
\label{eq:eigen}
  \myshifted{\xi}{ e^{-i \phi(\boldsymbol{z})} } 
  = 
  e^{-i \phi(\boldsymbol{z})} / ( 1 - c\xi ).
\end{equation} 
Considering our problem, we, as in \cite{V15}, \emph{define}   
\begin{equation}
\label{sol:shab}
  \mathbb{E}_{\xi} \langle a | = \langle a |  \mathsf{J}_{\xi}^{-1}, \qquad 
  \mathbb{E}_{\xi} \langle b | = \langle b |  \mathsf{K}_{\xi}
\end{equation}
with 
\begin{equation}
  \mathsf{J}_{\xi} = 1 - \xi  \mathsf{R}^{-1}, \qquad
  \mathsf{K}_{\xi} = 1 - \xi  \mathsf{L}^{-1} 
\end{equation}
(a vector version of \eref{eq:eigen}) 
which clearly implies 
\begin{equation}
\label{sol:shAB}
  \mathbb{E}_{\xi} \mathsf{A} = \mathsf{A} \mathsf{J}_{\xi}^{-1}, \qquad 
  \mathbb{E}_{\xi} \mathsf{B} = \mathsf{B} \mathsf{K}_{\xi}
\end{equation}
and then {\emph{prove} in Appendix B that 
functions \eref{sol:q1}--\eref{sol:q4} satisfy 
\eref{shx:q1} and \eref{shx:q4}.
Moreover, these functions are solutions of all 
differential equations of the ALH
(equations \eref{dx:q1}--\eref{dt:q4} included), 
provided the variables of the hierarchy, 
$z_{1},z_{2},...$, are introduced in accordance with the definition 
\eref{sol:shab}. 
Returning to our problem, this means that the $(\z1,\z3)$-dependency is 
governed by 
\begin{equation}
  i \mathsf{A}_{\z1} = \mathsf{A}  \mathsf{R}^{-1}, \qquad 
  i \mathsf{A}_{\z3} = \mathsf{A} \mathsf{R}^{-2}
\end{equation}
and 
\begin{equation}
  i \mathsf{B}_{\z1} = -  \mathsf{B}  \mathsf{L}^{-1}, \qquad 
  i \mathsf{B}_{\z3} = -  \mathsf{B}  \mathsf{L}^{-2},
\end{equation}
with similar equations for $\langle a |$ and $\langle b |$. 
Summarizing, we can formulate the following result.

\begin{proposition} \label{prop1}

Vectors $\myvec{q}$ and $\myvec{r}$ defined in \eref{def:qr} and 
\eref{sol:q1}--\eref{sol:q4} with
\begin{equation}
  \langle a(\z1,\z3) | = \langle a_{0} | \mathsf{E}_{A}(\z1,\z3), \quad 
  \mathsf{A}(\z1,\z3) = \mathsf{A}_{0} \mathsf{E}_{A}(\z1,\z3), 
\end{equation}
\begin{equation}
  \langle b(\z1,\z3) | = \langle b_{0} | \mathsf{E}_{B}(\z1,\z3), \quad 
  \mathsf{B}(\z1,\z3) = \mathsf{B}_{0} \mathsf{E}_{B}(\z1,\z3), 
\end{equation}
where 
$\langle a_{0} | = \left( a_{01}, ... , a_{0N} \right)$ 
and
$\langle b_{0} | = \left( b_{01}, ... , b_{0N} \right)$ 
are arbitrary constant $N$-rows, 
\begin{equation}
\label{def:AB0}
  \mathsf{A}_{0} = 
  \Biggl( \frac{ \alpha_{j}a_{0k} }{ L_{j} - R_{k} } \Biggr)_{j,k = 1, ..., N}, 
  \qquad
  \mathsf{B}_{0} = 
  \Biggl( \frac{ \beta_{j}b_{0k} }{ R_{j} - L_{k} } \Biggr)_{j,k = 1, ...,N} 
\end{equation}
and
\begin{equation}
  \mathsf{E}_{A}(\z1,\z3) = 
  \exp\left( - i \z1 \mathsf{R}^{-1} - i \z3 \mathsf{R}^{-2} \right),
\end{equation}
\begin{equation}
  \mathsf{E}_{B}(\z1,\z3) = 
  \exp\left( i \z1 \mathsf{L}^{-1} + i \z3 \mathsf{L}^{-2} \right) 
\end{equation}
satisfy equations \eref{nlse:q}--\eref{eq:lambda} and constraint \eref{uni:qr}. 
\end{proposition}
Thus, we have derived soliton solutions for the `two-field' version of our 
problem, consisting of linear Schr\"{o}dinger equations under 
the bilinear constraint \eref{uni:qr}.

The last step is to pass from solutions described in Proposition \ref{prop1} 
to solution of our problem. We start with imposing some restrictions on the 
constants involved to ensure the relations
\begin{equation} 
  \q2 = \q1{\!}^{*}, \qquad \q4 = \q3{\!}^{*}.
\end{equation} 
It turns out that this can be achieved by taking 
\begin{equation} 
  R_{j} = L_{j}^{*}, 
  \quad
  \beta_{j} a_{0j} = (\alpha_{j}b_{0j})^{*},
  \qquad j=1, ..., N. 
\end{equation} 
This, at first, implies
$ \mathsf{E}_{B} = \mathsf{E}_{A}^{*}$. 
Then, after rewriting \eref{def:AB0} as
$\mathsf{A}_{0} = \mathsf{D}_{\alpha} \mathsf{C} \mathsf{D}_{a}$ and 
$\mathsf{B}_{0} = \mathsf{D}_{\beta}  \mathsf{C}^{*} \mathsf{D}_{b}$, 
where
\begin{equation} 
\label{def:C}
  \mathsf{C} = 
  \Biggl( \frac{ 1 }{ L_{j} - L_{k}^{*} } \Biggr)_{j,k = 1, ..., N} 
\end{equation} 
and
$\mathsf{D}_{\alpha}$, $\mathsf{D}_{\beta}$, $\mathsf{D}_{a}$, $\mathsf{D}_{b}$ 
are diagonal matrices with elements 
$\alpha_{j}$, $\beta_{j}$, $a_{0j}$, $b_{0j}$ ($j = 1, ..., N$) correspondingly,
one can present $\mathsf{A}$ and $\mathsf{B}$ as 
\begin{equation} 
  \mathsf{A} = \mathsf{D}_{\alpha} \mathsf{Y} \mathsf{D}_{\beta}^{-1}, \quad
  \mathsf{B} = \mathsf{D}_{\beta} \mathsf{Y}^{*} \mathsf{D}_{\alpha}^{-1} 
\end{equation} 
where 
\begin{equation}
\label{def:Y} 
  \mathsf{Y} = \mathsf{C} \mathsf{D}_{\beta}\mathsf{D}_{a} \mathsf{E}_{A}.
\end{equation} 
which leads to 
\begin{equation} 
  \mathsf{F} = 
    \mathsf{D}_{\beta} 
    \left( \mathsf{1} + \mathsf{Y}^{*} \mathsf{Y} \right)^{-1} 
    \mathsf{D}_{\beta}^{-1},
  \qquad
  \mathsf{G} = 
    \mathsf{D}_{\alpha} 
    \left( \mathsf{1} + \mathsf{Y} \mathsf{Y}^{*} \right)^{-1} 
    \mathsf{D}_{\alpha}^{-1}.
\end{equation} 
After some simple calculations, one can rewrite $\q1$ and $\q3$ as
\begin{equation} 
\label{sol:q13}
  \q1 = 
    \langle \gamma | \mathsf{\Omega} | 1 \rangle, \quad
  \q3 = 1 + \langle \gamma | \mathsf{\Omega} \mathsf{Y}^{*} | 1 \rangle
\end{equation} 
where 
\begin{equation} 
\label{def:Omega}
    \mathsf{\Omega} = 
    \mathsf{Y} 
    \left( \mathsf{1} + \mathsf{Y}^{*} \mathsf{Y} \right)^{-1}, 
\end{equation} 
the constant row $\langle \gamma |$ is given by 
\begin{equation} 
\label{def:gamma}
  \langle \gamma | 
  = 
  \langle 1 | \left( \mathsf{C}\mathsf{L}^{*} \right)^{-1} 
\end{equation} 
and
\begin{equation} 
  \langle 1 | = \left( 1, ..., 1 \right), \quad
  | 1 \rangle = \left( 1, ..., 1 \right)^{T}. 
\end{equation} 

In a similar way, the ansatz for $\q2$ and $\q4$,
given by \eref{sol:q2} and \eref{sol:q4}, rewritten in terms of 
$\mathsf{Y}$, $\mathsf{\Omega}$ and $\langle \gamma |$ leads to 
\begin{equation} 
  \q2 = 
    \langle \gamma^{*} | \mathsf{\Omega}^{*} | 1 \rangle, \quad
  \q4 = 1 + \langle \gamma^{*} | \mathsf{\Omega}^{*} \mathsf{Y} | 1 \rangle
\end{equation} 
which demonstrates that $\q2 = \q1\strut^{\!*}$ and $\q4 = \q3\strut^{\!*}$.

The dependence on $t$ and $x$ is described by the matrix $\mathsf{E}$, 
\begin{equation}
  \mathsf{E} = \mathsf{D}_{\beta}\mathsf{D}_{a} \mathsf{E}_{A} 
\end{equation} 
which can be written as 
\begin{equation} 
\label{def:E}
  \mathsf{E} = 
  \mathop{\mbox{diag}}\left( \; e^{f_{k} + i\varphi_{k}} \; \right)_{k=1, ..., N}
\end{equation} 
with 
\begin{equation} 
  \begin{array}{lcl} 
  f_{k}(t,x)   &=& \nu_{k} x  + 2\mu_{k}\nu_{k} t + f_{0k}, \\[2mm]
  \varphi_{k}(t,x) &=& - \mu_{k} x  + (\nu_{k}^{2}-\mu_{k}^{2}) t + \varphi_{0k} 
  \end{array} 
\end{equation} 
where $\mu_{k}$ and $\nu_{k}$ are defined by 
\begin{equation} 
  \mu_{k} + i\nu_{k} = L_{k} \left/ |L_{k}|^{2} \right.
\end{equation} 
and $f_{0k}$, $\varphi_{0k}$ are arbitrary constants.

Now, we have all necessary to formulate the main result of this paper.

\begin{proposition} \label{prop2}

Vectors $\myvec{\psi}$ defined by 
\begin{equation} 
  \myvec{\psi} = 
  \mathsf{U}
  \left( \begin{array}{c} 
    \langle \gamma | \mathsf{\Omega} | 1 \rangle, \quad
  \\  
    1 + 
    \langle \gamma | \mathsf{\Omega} \mathsf{Y}^{*} 
    | 1 \rangle
  \end{array}  \right)
\end{equation} 
where $\mathsf{U}$ is an arbitrary constant unitary matrix,
\begin{equation} 
  \mathsf{\Omega} = 
    \mathsf{Y} 
    \left( \mathsf{1} + \mathsf{Y}^{*} \mathsf{Y} \right)^{-1}, \quad 
  \mathsf{Y} = \mathsf{C} \mathsf{E} , 
\end{equation} 
with $\langle \gamma |$, $\mathsf{C}$ and $\mathsf{E}$
defined in \eref{def:gamma}, \eref{def:C} and \eref{def:E}, 
solve the Euler–Lagrange equations \eref{eq:snlse}. 
Elements of the matrix $\mathsf{U}$ together with the 
$4N$ constants $\mathrm{Re}\,L_{k}$, $\mathrm{Im}\,L_{k}$, 
$f_{0k}$ and $\varphi_{0k}$ are 
arbitrary parameters that determine the properties of the $N$-soliton 
solution.
\end{proposition}

To get some insight into the structure of obtained solutions, let us consider 
the one-soliton solution.

In this case the matrices $\mathsf{L}$, $\mathsf{C}$ and 
the rows $\langle \gamma |$ become just scalars,
\begin{equation} 
  \mathsf{L} \to L_{1}, \quad 
  \mathsf{C} \to \frac{1}{2 i |L_{1}| \sin\theta}, \quad 
  \langle \gamma | \to e^{2 i \theta} - 1
\end{equation} 
where $\theta = \arg L_{1}$, or 
\begin{equation} 
  \theta = \tan^{-1}(\nu_{1}/\mu_{1}). 
\end{equation} 
Modifying slightly the functions $f_{1}$ and $\varphi_{1}$ that appear in 
\eref{def:E}, i.e. introducing the new ones, $f$ and $\varphi$, defined by 
\begin{equation} 
  f(t,x) = 
  f_{1}(t,x) 
  + \ln \frac{\mu_{1}^{2} + \nu_{1}^{2}}{ 2 |\nu_{1}|}, 
\end{equation} 
\begin{equation} 
  \varphi(t,x) = \varphi_{1}(t,x) - \frac{\pi}{2} \mathop{\mbox{sign}}\nu_{1} 
\end{equation} 
one can present $\mathsf{Y}$ as $\mathsf{\Omega}$ as
\begin{equation} 
  \mathsf{Y} = \exp( f + i\varphi ), \quad
  \mathsf{\Omega} = \frac{1}{2} \exp( i\varphi )\mathop{\mbox{sech}}f. 
\end{equation} 
Choosing 
$\mathsf{U} = \mbox{diag}\left( e^{-i\theta},e^{-i\theta} \right)$ 
one can obtain the following expressions for the components of the vector 
$\myvec{\psi} = \left( \psi_{1}, \psi_{2} \right)^{\smallT}$:
\begin{equation} 
\label{equ:psi}
  \psi_{1} = i \sin\theta \, \exp(i\varphi) \mathop{\mbox{sech}}{f}, 
\end{equation} 
\begin{equation} 
  \psi_{2} = \cos\theta + i \sin\theta \, \tanh f. 
\end{equation} 
with 
\begin{equation} 
  f(t,x) = \nu_{1} x  + 2\mu_{1}\nu_{1} t + f_{0},
\end{equation} 
\begin{equation} 
\label{equ:phi}
  \varphi(t,x) = - \mu_{1} x  + (\nu_{1}^{2}-\mu_{1}^{2}) t + \varphi_{0}. 
\end{equation} 

It is easy to see that $\psi_{1}$ and $\psi_{2}$ are the NLSE-solitons of 
different type: $\psi_{1}$ is a so-called bright soliton, vanishing as 
$\z1 \to \pm\infty$ (with $\z3$ being fixed), 
while $\psi_{2}$ is a dark soliton 
($\lim_{|\z1| \to \infty} \psi_{2} = e^{ \pm i \theta}$). 
However, this does not mean that the same is 
true for any solution (even for the one-soliton one), because for an 
arbitrary $\mathsf{U}$, all components of 
the general $N$-soliton solution are mixtures of dark and bright solitons.

Another useful information that can be obtained form 
\eref{equ:psi}--\eref{equ:phi} is the fact that all physically important 
characteristics of the soliton, its amplitude ($=\sin\theta$), velocity 
($=-2\mu_{1}$) and the scale ($=1/\nu_{1}$), are determined by the choice of 
$L_{1}$. 
The same is true in the multi-soliton case. 
Although a $N$-soliton solution of a \emph{nonlinear} equation is surely not a 
sum of $N$ solitons (only in the asymptotic regions, $t \to \pm\infty$ it can 
be viewed as such), it is possible to describe its structure qualitatively in 
terms of single solitons whose amplitude, velocity and scale depend on on the 
elements of the matrix $\mathsf{L}$, $L_{k}$, while other parameters, 
$\alpha_{k}$, $\beta_{k}$, $a_{0k}$ and $b_{0k}$, 
which were `absorbed' into the matrix $\mathsf{E}$ \eref{def:E}, 
i.e. replaced by $f_{0k}$ and $\varphi_{0k}$, 
determine their relative position and phasing. 
The role of the matrix $\mathsf{U}$ is mostly `mixing' of different solitons.

\section{Vector NLSE with gradient nonlinearity. \label{sec:gnlse} }

In this section we consider an example of application of the results, 
obtained in this paper for a rather abstract system, 
to a more physical problem.

Let us return to the system \eref{nlse:q}, \eref{nlse:r} and study the 
behavior of the function $s$ defined as
\begin{equation}
\label{def:s}
 s = \mycev{r}\myvec{q}_{\z1} = - \mycev{r}_{\z1}\myvec{q}
\end{equation}
(the last equality stems from the fact 
$\partial_{\z1} (\mycev{r}\myvec{q}) = 0$).
Calculating the derivative of $s$ using both of the equations in \eref{def:s} 
and expressing 
$\myvec{q}_{\z1\z1}$ and $\mycev{r}_{\z1\z1}$ from \eref{nlse:q} and \eref{nlse:r}, 
one can easily obtain
\begin{eqnarray} 
  s_{\z1} 
  & = & 
    - i \mycev{r}\myvec{q}_{\z3} + \mycev{r}_{\z1}\myvec{q}_{\z1} - \lambda \\
  & = & 
    - i \mycev{r}_{\z3}\myvec{q} - \mycev{r}_{\z1}\myvec{q}_{\z1} + \lambda             
\end{eqnarray} 
which, after summation of both expressions, leads to 
\begin{equation} 
  s_{\z1} = - (i/2) \, \partial_{\z3} (\mycev{r}\myvec{q}) = 0 
\end{equation} 
which means that $s = s(t)$.

In a similar way, one can calculate the time derivative of $s$ which leads to 
\begin{equation} 
  i s_{\z3} = 
  \partial_{\z1}\left( 2\mycev{r}_{\z1}\myvec{q}_{\z1} - \lambda \right).
\end{equation} 
This implies that
\begin{equation} 
  \lambda = \lambda_{0} + 2\mycev{r}_{\z1}\myvec{q}_{\z1}  
\end{equation} 
where 
\begin{equation} 
  \lambda_{0} = - i s'(\z3)\z1 + s_{1}(\z3) 
\end{equation} 
and $s_{1}$ is another functions not depending on $\z1$.
Thus, equations \eref{nlse:q} and \eref{nlse:r} can be rewritten as 
\begin{equation}
  i \myvec{q}_{\z3} 
  + \myvec{q}_{\z1\z1} 
  + (\lambda_{0} + 2\mycev{r}_{\z1}\myvec{q}_{\z1}) \myvec{q}
  = 0,
\end{equation}
\begin{equation}
  - i \mycev{r}_{\z3} 
  + \mycev{r}_{\z1\z1} 
  + (\lambda_{0} + 2\mycev{r}_{\z1}\myvec{q}_{\z1}) \mycev{r}
  = 0
\end{equation}

In the case of the soliton solutions presented in the previous section, 
it can be shown, by calculating $\mycev{r}_{\z1}\myvec{q}_{\z1}$
using equations \eref{eq:dxq} and comparing the result with the expression 
for $\lambda$ in \eref{lambdauv}, that $\lambda_{0} = 0$.
Thus, as a byproduct of the calculations presented in this paper, we obtain 
the following result.

\begin{proposition}
$N$-soliton solutions described in proposition \ref{prop2} are, at the same 
time, solutions of the NLSE with gradient nonlinearity
\begin{equation} 
  i \myvec{\psi}_{t} 
  + \myvec{\psi}_{xx} 
  + 2 \, \myvec{\psi}^{\dagger}_{x} \myvec{\psi}_{x} \, \myvec{\psi} 
  = 0.
\end{equation} 

\end{proposition}

\section{Conclusion.}

In this work we have established the relationship between the 
Schr\"{o}dinger equation with the constraint and the ALH. As was mentioned in 
section \ref{sec:ALH}, we considered only the `positive' subhierarchy 
\eref{alh:tau}. As to the `negative' subhierarchy, it can be shown that 
calculations similar to ones presented above lead to the set of solutions 
similar to the solutions described in the proposition \ref{prop2}.

A more interesting question is whether we can tackle with the approach of this 
paper the case on quadratic restrictions other than \eref{constraint}, for 
example, ones given by
\begin{equation} 
  \myvec{\psi}^{\dagger} \sigma_{3}\myvec{\psi} = 1
\end{equation} 
where $\sigma_{3} = \mbox{diag}\left(1,-1\right)$? 
The answer, which we present here without derivation, is `yes'. However, to do 
this one should start not with the bright solitons of the ALH, as in section 
\ref{sec:solitons}, but with the \emph{dark} ones. 
In some sense the signature of the matrix describing the applied 
constraints play the role of the sign in front of the nonlinear term in the 
NLSE: it determines which kind solitons 
(bright or dark) exists in the system.

Finally we would like to add a short comment on the integrability of the 
system \eref{eq:snlse} which was not discussed in the paper. We cannot at 
present prove its integrability, for example, by developing the inverse 
scattering transform. 
However, we now know that equations \eref{eq:snlse} possess $N$-soliton 
solutions. That means that the model \eref{Lagrangian} passes the so-called 
$N$-soliton test for integrability, which, though not proved rigorously, has a 
long history of successful applications and was even used as a tool for finding 
new integrable models, as, for example, in the comprehensive study by Hietarinta 
\cite{H87a,H87b,H87c,H88}.
Also, we now know that equations \eref{eq:snlse} are a consequence of 
equations of the integrable ALH.
These two facts are a strong indication that the model \eref{Lagrangian} is 
integrable. 
Nevertheless, we think that the work in this direction should be continued, 
and one of the first problems to solve is to find the conservation laws of the 
model, which may be a topic of the following study.

\section*{Acknowledgments.}

We would like to thank the Referees for 
careful reading of the manuscript and their valuable comments and suggestions, 
which helped to improve this paper.

\section*{Appendix A. A proof of \eref{nlse:uv}.}

\setcounter{equation}{0}
\def\theequation{A.\arabic{equation}}

To prove the fact that functions $\q5$ and $\q6$ satisfy the NLSE we need 
some consequences of the formulae presented in section \ref{sec:ALH}.

First, it follows from equations \eref{dx:q1} and \eref{dx:q4} together with 
the restriction \eref{uni:q}, that functions $\q5$ and $\q6$ can be written as
\begin{equation}
\label{A:1}
   \q5 = i \q1_{\z1}\q3 - i \q1\q3_{\z1}, 
\qquad 
   \q6 = i \q2\q4_{\z1} - i \q2_{\z1}\q4 
\end{equation}
and that 
\begin{equation}
\label{A:2}
   \q5\q6 = \q1_{\z1}\q2_{\z1} + \q3_{\z1}\q4_{\z1}. 
\end{equation}
Differentiating \eref{A:1} with respect to $\z3$ and expressing the 
$\z3$-derivatives from \eref{dt:q1} and \eref{dt:q4} leads to
\begin{equation}
\label{A:3}
  \begin{array}{lcl} 
  \q5_{\z3} & = &
  - \q5_{\z1\z1} + \left( \q1_{\z1}\q2 + \q3_{\z1}\q4 \right) \q5_{\z1} 
  \\&&  
  + \left( 
    \q1\q2_{\z1\z1} + \q3\q4_{\z1\z1} 
    - \q1_{\z1}\q2_{\z1} - \q3_{\z1}\q4_{\z1}
    \right) \q5. 
  \end{array}
\end{equation}
Another consequence of \eref{dx:q1} and \eref{dx:q4} is that 
\begin{equation} 
  \q1_{\z1}\q2 + \q3_{\z1}\q4 = 0, \qquad
  \q1\q2_{\z1} + \q3\q4_{\z1} = 0
\end{equation} 
which, in particular, implies 
\begin{equation} 
  \q1\q2_{\z1\z1} + \q3\q4_{\z1\z1} = 
  - \q1_{\z1}\q2_{\z1} - \q3_{\z1}\q4_{\z1}.
\end{equation} 
Thus, the first braces in the right-hand side of equation \eref{A:3} disappear 
while the factor in front of $\q5$ becomes just $-2\q5\q6$. Equation \eref{A:3} 
now reads
\begin{equation}
  \q5_{\z3} = - \q5_{\z1\z1} - 2 \q5^{2}\q6 
\end{equation}
which is nothing but the first equation from \eref{nlse:uv}.

The second equation from \eref{nlse:uv} can be demonstrated in the similar way.

\section*{Appendix B. Validation of the \textit{ansatz}.} 

\setcounter{equation}{0}
\def\theequation{B.\arabic{equation}}

Here we demonstrate that \textit{ansatz} \eref{sol:q1}--\eref{sol:q4} 
leads to the solutions of the \eref{shx:q1} and \eref{shx:q4}

As follows from \eref{sol:q1} and \eref{sol:shab}, 
\begin{eqnarray}
  \myshifted{\xi}{\q1} - \q1
  & = & 
  \langle\myshifted{\xi}{{a}}|  
    \mathsf{R}^{-1} (\myshifted{\xi}{ \mathsf{F}}) 
  | \beta \rangle
  - 
  \langle{a}| 
    \mathsf{R}^{-1} \mathsf{F} 
  | \beta \rangle
\nonumber\\
  & = & 
  \langle\myshifted{\xi}{{a}}| 
    \mathsf{R}^{-1} (\myshifted{\xi}{ \mathsf{F}}) 
  | \beta \rangle
  - 
  \langle\myshifted{\xi}{{a}}| 
    \mathsf{R}^{-1}  \mathsf{J}_{\xi}  \mathsf{F} 
  | \beta \rangle
\end{eqnarray}
which can be rewritten as 
\begin{equation}
\label{B:1}
  \myshifted{\xi}{\q1} - \q1
  = 
  \langle\myshifted{\xi}{{a}}|  
    \mathsf{R}^{-1} (\myshifted{\xi}{ \mathsf{F}}) \mathsf{X} \mathsf{F} 
  | \beta \rangle
\end{equation}
where $\mathsf{X}$ is defined by 
\begin{equation}
  (\myshifted{\xi}{ \mathsf{F}}) \mathsf{X} \mathsf{F}
  = 
  \myshifted{\xi}{ \mathsf{F}} -  \mathsf{J}_{\xi}  \mathsf{F}.
\end{equation}
Using the definition of $\mathsf{F}$ \eref{sol:FG}, \eref{sol:shAB} and then 
\eref{sol:sy} one can obtain 
\begin{eqnarray}
  \mathsf{X}
  & = & 
  1 - \mathsf{J}_{\xi}
  + \mathsf{B}  \mathsf{A}
  - (\myshifted{\xi}{ \mathsf{B}}) 
    (\myshifted{\xi}{ \mathsf{A}}) 
    \mathsf{J}_{\xi}
\nonumber\\
  & = & 
  \xi  \mathsf{R}^{-1}
  + \mathsf{B} \mathsf{A}
  - (\myshifted{\xi}{ \mathsf{B}})  \mathsf{A}
\nonumber\\
  & = & 
  \xi  \mathsf{R}^{-1}
  + \xi  \mathsf{B}  \mathsf{L}^{-1}  \mathsf{A}
\nonumber\\
  & = & 
  \xi  \mathsf{R}^{-1}  \mathsf{F}^{-1}
  + \xi \mathsf{R}^{-1} | \beta \rangle \langle b |  \mathsf{L}^{-1}  \mathsf{A}. 
\end{eqnarray}
Substituting this expression into \eref{B:1} and using \eref{sol:q4} together 
with the identity $\mathsf{A}\mathsf{F} = \mathsf{G}\mathsf{A}$, one arrives at 
\begin{eqnarray}
  \myshifted{\xi}{\q1} - \q1
  & = & 
  \xi \langle\myshifted{\xi}{{a}}| 
    \mathsf{R}^{-1} (\myshifted{\xi}{ \mathsf{F}}) \mathsf{R}^{-1} 
  | \beta \rangle
+ \xi \langle\myshifted{\xi}{{a}}| 
    \mathsf{R}^{-1} (\myshifted{\xi}{ \mathsf{F}}) \mathsf{R}^{-1} 
  | \beta \rangle 
  \langle b | 
    \mathsf{L}^{-1} \mathsf{A} \mathsf{F} 
  | \beta \rangle
\nonumber\\
  & = & 
 \xi \q4 \myshifted{\xi}{{\q5}}
\end{eqnarray}
which is nothing but the first equation from \eref{shx:q1}.

In a similar way one can prove that $\q3$ satisfies the first equation from 
\eref{shx:q4}.
\begin{eqnarray}
  \myshifted{\xi}{\q3} - \q3
  & = & 
  \langle\myshifted{\xi}{{a}}| 
    \mathsf{R}^{-1}(\myshifted{\xi}{\mathsf{F}})(\myshifted{\xi}{\mathsf{B}}) 
  | \alpha \rangle
  - 
  \langle a |  \mathsf{R}^{-1}  \mathsf{F}  \mathsf{B} | \alpha \rangle
\nonumber \\
\label{B:2}
  & = & 
  \langle\myshifted{\xi}{{a}}| 
    \mathsf{R}^{-1} (\myshifted{\xi}{ \mathsf{F}}) \mathsf{Y} \mathsf{G} 
  | \alpha \rangle
\end{eqnarray}
where 
\begin{equation}
  (\myshifted{\xi}{ \mathsf{F}}) \mathsf{Y} \mathsf{G}
  = 
  (\myshifted{\xi}{ \mathsf{F}}) (\myshifted{\xi}{ \mathsf{B}}) 
  -  \mathsf{J}_{\xi}  \mathsf{F}  \mathsf{B}.
\end{equation}
Calculating $\mathsf{Y}$,
\begin{eqnarray}
  \mathsf{Y}
  & = & 
  \myshifted{\xi}{ \mathsf{B}}
  - \mathsf{J}_{\xi}  \mathsf{B}
  + (\myshifted{\xi}{ \mathsf{B}}) \mathsf{A} \mathsf{B}
  - (\myshifted{\xi}{ \mathsf{B}}) (\myshifted{\xi}{ \mathsf{A}}) \mathsf{J}_{\xi}  \mathsf{B}
\nonumber\\
  & = & 
  \myshifted{\xi}{ \mathsf{B}} -  \mathsf{J}_{\xi}  \mathsf{B}
\nonumber\\
  & = & 
  \mathsf{B}  \mathsf{K}_{\xi} -  \mathsf{J}_{\xi}  \mathsf{B}
\nonumber\\
  & = & 
  \xi  \mathsf{R}^{-1}  \mathsf{B} - \xi  \mathsf{B}  \mathsf{L}^{-1}
\nonumber\\
  & = & 
  - \xi  \mathsf{R}^{-1} | \beta \rangle \langle b |  \mathsf{L}^{-1},
\end{eqnarray}
and substituting it in \eref{B:2}
one can obtain 
\begin{eqnarray}
  \myshifted{\xi}{\q3} - \q3
  & = & 
  - \xi \langle b |
      \mathsf{L}^{-1}  \mathsf{G} 
    | \alpha \rangle 
  \langle\myshifted{\xi}{{a}}| 
    \mathsf{R}^{-1} (\myshifted{\xi}{ \mathsf{F}})  \mathsf{R}^{-1} 
  | \beta \rangle
\nonumber\\
  & = & 
- \xi \q2 \myshifted{\xi}{{\q5}},
\end{eqnarray}
which concludes the proof.

The rest of the equations \eref{shx:q1} and \eref{shx:q4} can be tackled in a 
similar way.


\begin{thebibliography}{99} 

\bibitem{P76} 
  Pohlmeyer K., 
  1976, 
  \href{https://doi.org/10.1007/BF01609119} 
       {\textit{Commun. Math. Phys.} \textbf{46}, 207--221}. 

\bibitem{K00} 
  Ketov S.V., 
  2000,
  \textit{Quantum Non-Linear Sigma-Models}
  (Springer-Verlag)
  (doi: 10.1007/978-3-662-04192-5).

\bibitem{F13} 
  Fradkin E., 
  2013, 
  \textit{Field Theories of Condensed Matter Physics}
  (Cambridge University Press).

\bibitem{V94} 
  Vekslerchik V.E., 
  1994, 
  \href{https://doi.org/10.1088/0305-4470/27/18/036} 
       {\textit{J. Phys. A} \textbf{27}, 6299--6313}. 

\bibitem{V20} 
  Vekslerchik V.E., 
  2020,
  \href{https://doi.org/10.3842/SIGMA.2020.144} 
       {\textit{SIGMA}, \textbf{16}, 144}. 

\bibitem{AL76} 
  Ablowitz M.J. and Ladik J.F., 
  1976,
  \href{https://doi.org/10.1063/1.523009} 
       {\textit{J. Math. Phys.} \textbf{17}, 1011--1018}. 

\bibitem{HJN16} 
  Hietarinta J., Joshi N., Nijhoff F. W., 
  2016,
  \textit{Discrete Systems and Integrability.} 
  (Cambridge University Press)
  (doi: 10.1017/CBO9781107337411). 

\bibitem{V15} 
  Vekslerchik V.E., 
  2015, 
  \href{https://doi.org/10.1088/1751-8113/48/44/445204} 
       {\textit{J. Phys. A} \textbf{48}, 445204}. 

\bibitem{H87a} 
  Hietarinta J., 
  1987, 
  \href{http://dx.doi.org/10.1063/1.527815}
       {\textit{J. Math. Phys.} {\bf 28}, 1732--1742}. 

\bibitem{H87b} 
  Hietarinta J., 
  1987, 
  \href{http://dx.doi.org/10.1063/1.527421}
       {\textit{J. Math. Phys.} {\bf 28}, 2094--2101}. 

\bibitem{H87c} 
  Hietarinta J., 
  1987, 
  \href{http://dx.doi.org/10.1063/1.527750}
       {\textit{J. Math. Phys.} {\bf 28}, 2586--2592}. 

\bibitem{H88} 
  Hietarinta J., 
  1988, 
  \href{http://dx.doi.org/10.1063/1.528002}
       {\textit{J. Math. Phys.} {\bf 29}, 628--635}. 

\end{thebibliography}
\end{document}